\documentclass[aip,jap,preprint,amsmath,amssymb,endfloats*]{revtex4-1}

\usepackage{graphicx}
\usepackage{dcolumn}
\usepackage{bm}
\usepackage{subfig}
\usepackage{caption}
\begin{document}

\title{An analysis method for asymmetric resonator transmission applied to superconducting devices}

\author{M. S. Khalil}
\email{moe@lps.umd.edu}
\affiliation{ 
Laboratory for Physical Sciences, College Park, MD, 20740
}
\affiliation{ 
Center for Nanophysics and Advanced Materials, Department of Physics, University of Maryland, College Park, MD, 20742
}%
\author{M. J. A. Stoutimore}%
\affiliation{ 
Laboratory for Physical Sciences, College Park, MD, 20740
}
\affiliation{ 
Center for Nanophysics and Advanced Materials, Department of Physics, University of Maryland, College Park, MD, 20742
}
\author{F. C. Wellstood}
\affiliation{ 
Center for Nanophysics and Advanced Materials, Department of Physics, University of Maryland, College Park, MD, 20742
}
\affiliation{ 
Joint Quantum Institute, University of Maryland, College Park, MD, 20742
}
\author{K. D. Osborn}%
\affiliation{ 
Laboratory for Physical Sciences, College Park, MD, 20740
}
\date{\today}

\begin{abstract}
We examine the transmission through nonideal microwave resonant circuits. The general analytical resonance line shape is derived for both inductive and capacitive coupling with mismatched input and output transmission impedances, and it is found that for certain non-ideal conditions the line shape is asymmetric. We describe an analysis method for extracting an accurate internal quality factor ($Q_i$), the Diameter Correction Method (DCM), and compare it to the conventional method used for millikelvin resonator measurements, the $\phi$ Rotation Method ($\phi$RM). We analytically find that the $\phi$RM deterministically overestimates $Q_i$ when the asymmetry of the resonance line shape is high, and that this error is eliminated with the DCM. A consistent discrepancy between the two methods is observed when they are used to analyze both simulations from a numerical linear solver and data from asymmetric coplanar superconducting thin-film resonators.
\end{abstract}


\maketitle 

\section{
\label{sec:level1}Introduction}

Precise measurements of both loaded and internal quality factors of thin-film superconducting resonators are necessary for many applications, from astronomy photon detectors, \cite{Day2003} to materials analysis, \cite{Martinis2005,Paik2010} to qubit readout. \cite{Wallraff2004, Houck2008} Growing interest in superconducting quantum computing has recently motivated detailed measurement of several types of superconducting thin-film resonators at millikelvin temperatures. \cite{ Martinis2005,OConnell2008,Gao2008,Wang2009,Cicak2010,Paik2010,Khalil2011,Gladchenko2011, Weber2011} Unfortunately, non-ideal experimental setups can lead to an asymmetry in the resonance line shape, \cite{OConnell2008,Gao2008a,Paik2010,Wisbey2010,Khalil2011,Klein1996} corresponding to a rotation of the resonance circle, which complicates the interpretation of internal and external quality factors ($Q_i$ and $Q_e$). Several methods for extracting the $Q_i$ of a resonator exist for different experimental setups. \cite{ Khanna1983,Kajfez1984,Kajfez1995,Leong2002,Klein1996} However, these methods either require a single port reflective measurement \cite{Kajfez1984,Kajfez1995} (incompatible with most qubit measurements), full two-port data \cite{Khanna1983,Leong2002} (typically unavailable for millikelvin measurements), or identifying and fitting to a second coupled mode \cite{Klein1996} (a special case). In contrast, the most widely used technique for analyzing millikelvin resonator measurements, the $\phi$ rotation method ($\phi$RM), simply adds an empirical rotation of the resonance circle to extract the quality factor. \cite{Paik2010, Gao2008a,Wisbey2010}

In this article we show how asymmetry in the resonance line shape can arise from coupling the resonator to mismatched input and output transmission lines and non-negligible transmission line series inductance. Based on this understanding of the origin of the asymmetry, we derive the Diameter Correction Method (DCM), used in recent publications, \cite{Khalil2011,Gladchenko2011} and use it to extract $Q_i$. We compare this to the conventional analysis method, the $\phi$ rotation method ($\phi$RM) \cite{Paik2010, Gao2008a,Wisbey2010}, and show that there is a one-to-one mapping between the two methods but that the $\phi$RM systematically overestimates $Q_i$ by an analytically quantifiable amount. Note, we will not address fitting techniques here because a comprehensive quantitative comparison of fitting techniques has been made.\cite{ Petersan1998}



\section{Derivation of Asymmetric Resonance}

\begin{figure}
\includegraphics[scale=1]{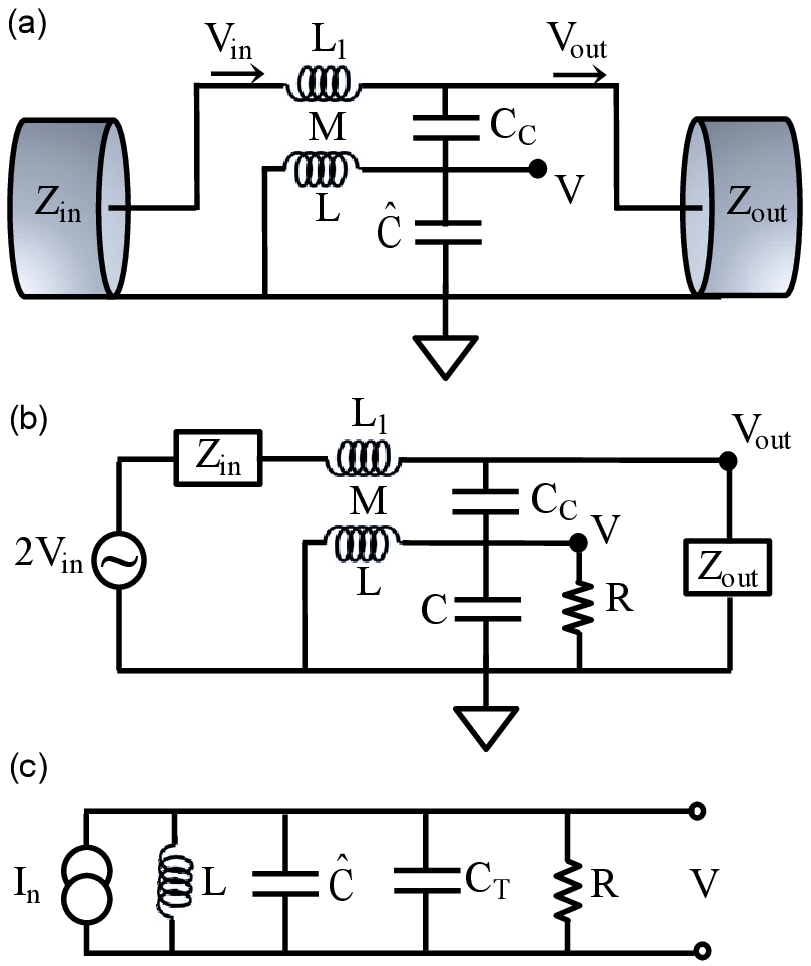}
\caption{\label{fig:schem} (a) Schematic of resonator measurement setup with both inductive and capacitive coupling and mismatched transmission lines. (b) equivalent circuit to (a), where $\hat{C}$ has been separated into its capacitive part (C) and  its resistive parts ($1/\omega R$). (c) Norton equivalent circuit for resonator measurement where V is the voltage across the capacitor, $\hat{C}$, and $G_N=R_T+1/\omega C_T$.}
\end{figure}

We consider a notch type resonator coupled to input and output transmission lines (see Fig.~\ref{fig:schem}(a)) in which the transmission, $S_{21}\equiv V_{out}/V_{in}$, is measured. Full transmission is measured off resonance and reduced transmission is measured on resonance. The resonator inductance and capacitance are $L$ and $\hat C$ , where $\hat C$ is complex to account for dielectric loss. $V_{in}$ and $V_{out}$ are the input and output voltage waves, and $V$ is the voltage across the capacitor, $\hat C$. Ideally, the transmission line ports are matched ($Z_ {in}=Z_{out}=Z_0$), and $L_1$ is small ($L_1<<C_CZ_{in}^2$ and  $L_1<<C_CZ_{out}^2$). However, conditions such as a precise matching of $Z_ {in}$ and $Z_ {out}$ can prove to be difficult to achieve in an experimental setup. 

Assuming a high quality factor, $Q_i>>1$, the circuit in Fig.~\ref{fig:schem}(a) can be redrawn as Fig.~\ref{fig:schem}(b), where $R=Q_i/\left(\omega_0 C\right)$. Solving Kirchhoff's equations we find an expression for the transmission as a function of the voltage across the capacitor:
\begin{align}
S_{21}=\left(1+\hat{\epsilon}\right)\left(1+\frac{V}{2V_{in}}\left(\frac{M}{L}+Z_{in}^\prime i\omega C_C\right)\right),
\label{eq:S21}
\end{align}
where $1+\hat{\epsilon}\equiv\frac{2}{1+\left(i\omega C_C+\frac{1}{Z_{out}}\right)Z_{in}^\prime}$, $Z_{in}^\prime \equiv Z_{in}+i\omega L_1-i\omega \frac{M^2}{L}$, and $|\hat{\epsilon}|<<1$. Solving Kirchhoff's equations again and eliminating $V_{out}$ we get the independent equation:
\begin{multline}
-V\Biggl(\frac{1}{i\omega L}+i\omega \hat{C}+\frac{i\omega C_C}{i \omega C_C Z_{out}+1}+\\
\frac{\left(\frac{M}{L}-\frac{i \omega C_C Z_{out}}{1+i \omega C_C Z_{out}}\right)^2}{Z_{out}^\prime+Z_{in}+i\omega \left(L_1-\frac{M^2}{L}\right)}\Biggr)=\\
2V_{in}\left(\frac{\frac{M}{L}-\frac{i \omega C_C Z_{out}}{1+i \omega C_C Z_{out}}}{Z_{out}^\prime+Z_{in}+i\omega \left(L_1-\frac{M^2}{L}\right)}\right),
\label{eq:Norton}
\end{multline}
where $Z_{out}^\prime \equiv \frac{Z_{out}}{1+i\omega C_CZ_{out}}$. Equation~(\ref{eq:Norton}) is of the form $V\left(\frac{1}{i\omega L} +i\omega \hat{C}+G_N\right)=I_N$, where $G_N$ and $I_N$ are the Norton equivalent conductance and current respectively. The circuit thus has a Norton equivalent shown in Fig.~\ref{fig:schem}(c). From Eq.~(\ref{eq:Norton}) we see that
\begin{align}
I_N=-2V_{in}\left(\frac{\frac{M}{L}-\frac{i \omega C_C Z_{out}}{1+i \omega C_C Z_{out}}}{Z_{out}^\prime+Z_{in}+i\omega \left(L_1-\frac{M^2}{L}\right)}\right),
\label{eq:In}
\end{align}
and,
\begin{align}
G_N=\frac{i\omega C_C}{i \omega C_C Z_{out}+1}+
\frac{\left(\frac{M}{L}-\frac{i \omega C_C Z_{out}}{1+i \omega C_C Z_{out}}\right)^2}{Z_{out}^\prime+Z_{in}+i\omega \left(L_1-\frac{M^2}{L}\right)}.
\label{eq:Gn}
\end{align}
Since the real part of $G_N$ loads the resonator measurement and the imaginary part shifts the resonance frequency, it is useful to separate $G_ N$ into its real (conductive) and imaginary (susceptive) components. We define $\mathbf{Re}\{G_N\} \equiv 1/R_T$, $\mathbf{Im}\{G_N\} \equiv \omega C_T$, $R _{eff} ^{-1} \equiv R^{-1}+R_T^{-1}$, $Q \equiv R_{eff}\omega_0(C+C_T)$, and $G ^\prime \equiv -\frac{I_N}{2V_{in}}\left(\frac{M}{L}+Z_{in}^\prime i \omega C_C\right)$. Then, Eq.~(\ref{eq:S21}) can be rewritten as
\begin{align}
S_{21}=\left(1+\hat{\epsilon}\right)\left(1-\frac{G ^\prime R_{eff}}{1+2iQ\frac{\omega-\omega_0}{\omega_0}}\right).
\label{eq:S21_G'}
\end{align}

We now note that to second-order in the small parameters $(M/L)$ and $\omega C_C Z_{out}$, $G ^\prime$ is equal to $R_T^{-1}$. Stopping at second-order would yield a resonance with a symmetric Lorentzian line shape. To expand to higher-order, we rewrite Eq.~(\ref{eq:S21_G'}) as 
\begin{align}
S_{21}=\left(1+\hat{\epsilon}\right)\left(1-\frac{\left(G_D+R_T^{-1}\right) R_{eff}}{1+2iQ\frac{\omega-\omega_0}{\omega_0}}\right),
\label{eq:S21_Gd}
\end{align}
where
\begin{align}
G_D \equiv G^\prime-R_T^{-1}
\label{eq:Gd}
\end{align}
Expanding to third-order we find:
\begin{multline}
G_D=i\omega C_C \frac{M}{L}\left(\frac{Z_{in}-Z_{out}}{Z_{in}+Z_{out}}\right)+\\
\frac{i}{\left(Z_{in}+Z_{out}\right)^2}\Biggl(\left(\omega C_C\right)^2Z_{out}^2\left(L_1-C_CZ_{in}^2\right)-\\
\left(\frac M L\right)^2\left(L_1-C_CZ_{out}^2\right)\Biggr),
\label{eq:Gd_expand}
\end{multline}
and note that $G_D$ is purely imaginary. We now define 
\begin{align}
\hat{Q}_e^{-1} \equiv \frac{R_T^{-1}+G_D}{\omega_0\left(C+C_T\right)},
\label{eq:Qe_complex}
\end{align}
and recognize that for $C_T<<C$
\begin{align}
Q_i^{-1}=Q^{-1}-\mathbf{Re}\left\{\hat{Q}_e^{-1}\right\},
\label{eq:Qi}
\end{align}
Thus we find
\begin{align}
S_{21}=\left(1+\hat{\epsilon}\right)\left(1-\frac{Q\hat{Q}_e^{-1}}{1+2iQ\frac{\omega-\omega_0}{\omega_0}}\right).
\label{eq:S21_complex_Qe}
\end{align}

\begin{figure}
\includegraphics[trim = 0 1.7cm 0 0, clip]{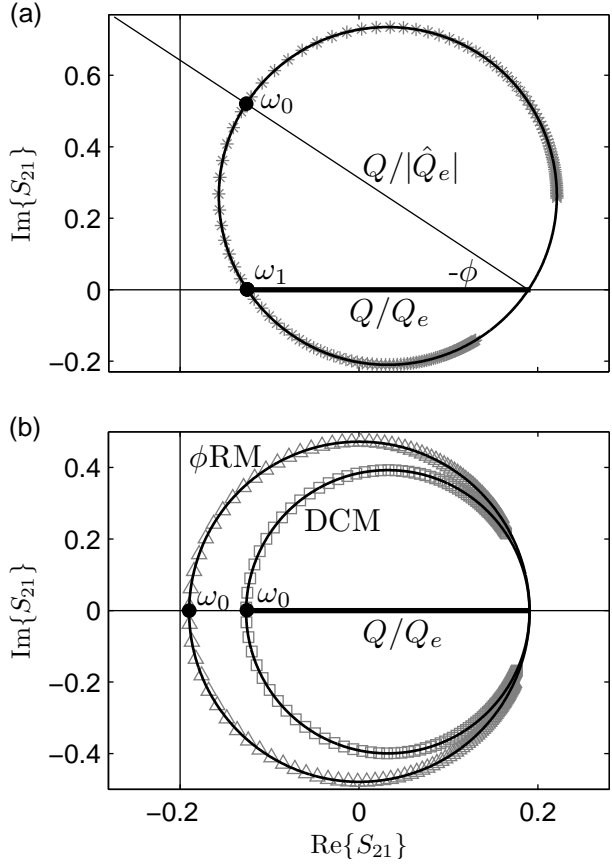}
\caption{\label{fig:circle} (a) Simulated transmission through mismatched coupling lines ($Z_{in}=24.5\Omega$, $Z_{out}=84.5 \Omega$) plotted as Im$\{S_{21}\}$ vs. Re$\{S_{21}\}$ with a fit to a circle. The asymmetry is represented as a rotation of the resonance circle by the angle $\phi$ away from the real (in-phase) axis or equivalently as $\delta\omega=\omega_1-\omega_0$, the frequency shift of the in-phase point on the resonance circle. (b) Shows the simulated transmission with asymmetry removed using both analysis techniques. The $\phi$RM ($\bigtriangleup$) only rotates the circle to the real axis while the DCM ($\square$) both rotates the circle and removes the factor of $1/\cos(\phi)$ increase to the diameter. The DCM shows that the invariant quantity is not, as the $\phi$RM assumes, the diameter of the circle (equal to $Q/|\hat{Q}_e|$ and $Q/Q_e$ before and after the DCM transformation respectively) but rather the length of the real axis segment intersecting the circle (shown in bold and equal to $Q/Q_e$), where $1/Q_e \equiv \mathbf{Re}\left\{1/\hat{Q}_e \right\}$.}
\end{figure}

We emphasize here that $G_D$ is the term that creates the asymmetry in the line shape or equivalently a rotation of the resonance circle around the off resonance point. If $G_D$ were zero then $\hat{Q}_e^{-1}$ would be real, reducing Eq.~(\ref{eq:S21_complex_Qe}) to a symmetric Lorentzian line shape. Equation~(\ref{eq:S21_complex_Qe}) can be written as
\begin{align}
S_{21}=\left(1+\hat{\epsilon}\right)\left(1-\frac{Q\left|\hat{Q}_e^{-1}\right|e^{i\phi}}{1+2iQ\frac{\omega-\omega_0}{\omega_0}}\right),
\label{eq:S21_phi}
\end{align}
where $\hat{Q}_e^{-1}$ is represented in terms of its magnitude and phase, $\phi$. Another equivalent representation is
\begin{align}
S_{21}=\left(1+\hat{\epsilon}\right)\left(1-\frac{\frac{Q}{Q_e}\left(1+2iQ\frac{\delta\omega}{\omega_0}\right)}{1+2iQ\frac{\omega-\omega_0}{\omega_0}}\right),
\label{eq:S21_delomega}
\end{align}
where we have defined $1/Q_e \equiv \mathbf{Re}\left\{\hat{Q}_e^{-1} \right\}$ and $\delta \omega$ is the difference between the resonance frequency and the new rotated in-phase point on the resonance circle, $\omega_1$ (see Fig.~\ref{fig:circle}). The form of Eq.~(\ref{eq:S21_delomega}) can be understood by noting that $S_{21}$ is real when $\omega=\delta \omega+\omega_0$ (to within a phase rotation of $1+\hat{\epsilon}$).

Here we stress that Eqs.~(\ref{eq:S21_complex_Qe}-\ref{eq:S21_delomega}) are equivalent representations of the asymmetric line shape, each highlighting a different interpretation of the asymmetry. In Eq.~(\ref{eq:S21_complex_Qe}) the asymmetry is quantified by $\mathbf{Im}\left\{\hat{Q}_e^{-1} \right\}$ and one can think of the asymmetry as coming from a complex loading of the resonator. In Eq.~(\ref{eq:S21_phi}) the asymmetry is quantified by $\phi$, where $\phi$ is the rotation angle of the resonance circle around the off-resonance point (see Fig.~\ref{fig:circle}). And finally in Eq.~(\ref{eq:S21_delomega}) the asymmetry is quantified by $\delta \omega$, where $\delta\omega=\omega_1-\omega_0$ is the frequency shift of the in-phase point on the resonance circle from $\omega_0$ to $\omega_1$ (see Fig.~\ref{fig:circle}). There of course exists a simple one-to-one mapping between the three notations:
\begin{align}
\phi=\arctan\left(\frac{\mathbf{Im}\{\hat{Q}_e^{-1}\}}{\mathbf{Re}\{\hat{Q}_e^{-1}\}}\right)=\arctan\left(2Q\frac{\delta\omega}{\omega_0}\right).
\label{eq:mapping}
\end{align}

One method that has been used \cite{Paik2010, Gao2008a,Wisbey2010} to extract internal quality factors simply accounts for the asymmetry by adding an empirical $\phi$ rotation and incorrectly substitutes $\left|\hat{Q}_e^{-1}\right|$ for $Q_e^{-1}$ by defining
\begin{align}
\frac{1}{Q_{i,\ \phi RM}}=\frac{1}{Q}-\left|\frac1{\hat{Q}_e}\right|.
\label{eq:Qi_wrong}
\end{align}
This method accounts for the asymmetric line shape phenomenologically by adding the rotation, $\phi$, without accounting for its origin and its impact on the interpretation of $Q_i$. It corresponds to rotating the resonance circle back an angle $\phi$, thereby putting $\omega_0$ on the in-phase axis. This is the $\phi$ rotation method ($\phi$RM), and the rotation of the $\phi$RM can best be seen by examining the difference between Fig.~\ref{fig:circle}(a) and Fig.~\ref{fig:circle}(b)($\bigtriangleup$). However, simply rotating the resonance circle by angle $\phi$ does not take into account the fact that the asymmetry has also caused the circle to grow by a factor of $1/\cos(\phi)$, assuming the circle has been normalized to full transmission off resonance $(S_{21}(\omega<<\omega_0)=S_{21}(\omega>>\omega_0)=1)$. We have shown here that instead one has
\begin{align}
\frac{1}{Q_{i,\ DCM}}=\frac{1}{Q}-\frac{1}{Q_e}.
\label{eq:Qi_right}
\end{align}
We call this the Diameter Correction Method (DCM) because in addition to rotating the circle by the asymmetry angle, $\phi$, it also corrects the diameter by accounting for the complex $Q_e$. This can be seen by examining the difference between Fig.~\ref{fig:circle}(a) and Fig.~\ref{fig:circle}(b)($\square$). Another interpretation of this result is that the quantity that remains constant in the asymmetry transformation is not the diameter of the resonance circle, as the $\phi$RM assumes, but rather the distance between the in-phase axis intercepts, shown in bold in Fig.~\ref{fig:circle}. That invariant length is the diameter of the circle for a symmetric resonance and becomes a chord of the circle as asymmetry is added, but remains equal to $Q/Q_e$, while the diameter grows as $Q/|\hat{Q}_e|$. The analytical discrepancy between the two methods can simply be determined by subtracting Eq.~(\ref{eq:Qi_right}) from Eq.~(\ref{eq:Qi_wrong}), 
\begin{align}
\frac{1}{Q_{i,\ DCM}}-\frac{1}{Q_{i,\ \phi RM}}=\left|\frac1{\hat{Q}_e}\right|\left(\cos(\phi)-1\right).
\label{eq:Qi_right_wrong_diff}
\end{align}
From Eq.~(\ref{eq:Qi_right_wrong_diff}) we see that the error in the $\phi$RM diverges for high asymmetry angle, $\phi \approx \pm \pi$, and for low $Q_e$, high coupling. Note that for $\phi=0$, $\hat{Q}_e^{-1}$ is real and Eq.~(\ref{eq:Qi_right}) reduces to Eq.~(\ref{eq:Qi_wrong}) and therefore Eq.~(\ref{eq:Qi_right_wrong_diff}) goes to zero.

\section{Fitting and Analyzing Simulations and Data}

To test the $\phi$RM and the DCM, we simulate the transmission using a numerical linear solver. Simulations are run varying several different parameters: $Q_i$, impedance mismatches, strength of both inductive and capacitive coupling and inductance of $L_1$. The resonator capacitance and inductance are held at 0.3 pF and 2.5 nH, respectively, producing a resonance frequency that ranges from 5.717-5.802 GHz (resonance frequency varies with coupling capacitance). The simulated data is then fit and analyzed using both methods.

\begin{figure}
\includegraphics[trim = 0.4cm 0.2cm 0.3cm 0, clip]{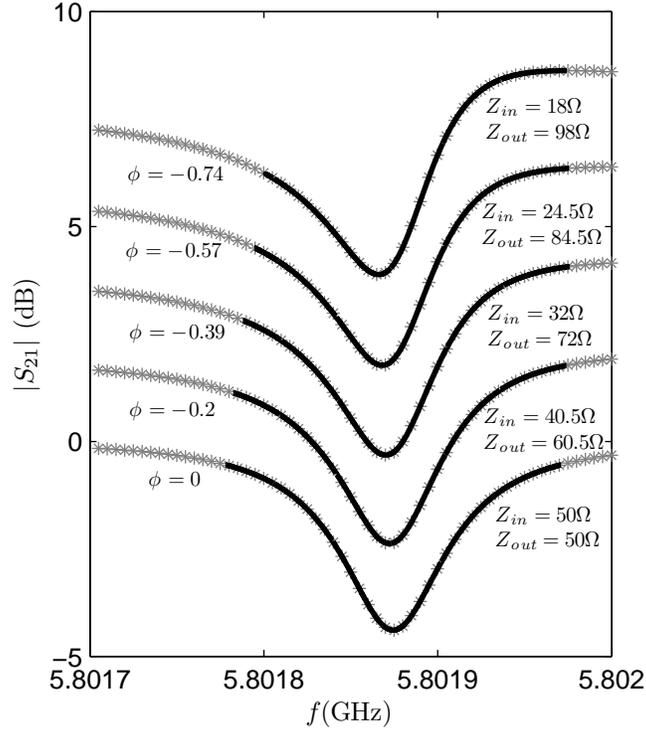}
\caption{\label{fig:lorent} Simulated and fit to symmetric and asymmetric resonance line shapes. Here the asymmetry is created using mismatched coupling lines ($Z_{in}$ and $Z_{out}$). Asymmetry angles, $\phi$, are extracted from the fits.}
\end{figure}

We created asymmetry by varying $Z_{in}/Z_{out}$. Note that asymmetry can also be created by increasing $L_1$. However, $L_1$ values in the nanohenries are required to create significant asymmetry which is far too large to be physical. Figure~\ref{fig:lorent} shows results from simulations and fits with the same simulated quality factor, $Q_i=10^5$ and a range of $Z_{in}/Z_{out}$ values. The coupling line mismatch creates a clear asymmetry in the line shape which is quantified  with the extracted asymmetry angle, $\phi$, also shown in Fig.~\ref{fig:lorent}. In addition to the value of $\phi$ extracted from the fit, we also analytically determine the asymmetry using
\begin{align}
\phi=\arctan\left(\frac{\mathbf{Im}\{G_D\}}{R_T^{-1}}\right).
\label{eq:phi}
\end{align}

\begin{figure}
\includegraphics[trim = 0.15cm 1.3cm 0.2cm 0, clip]{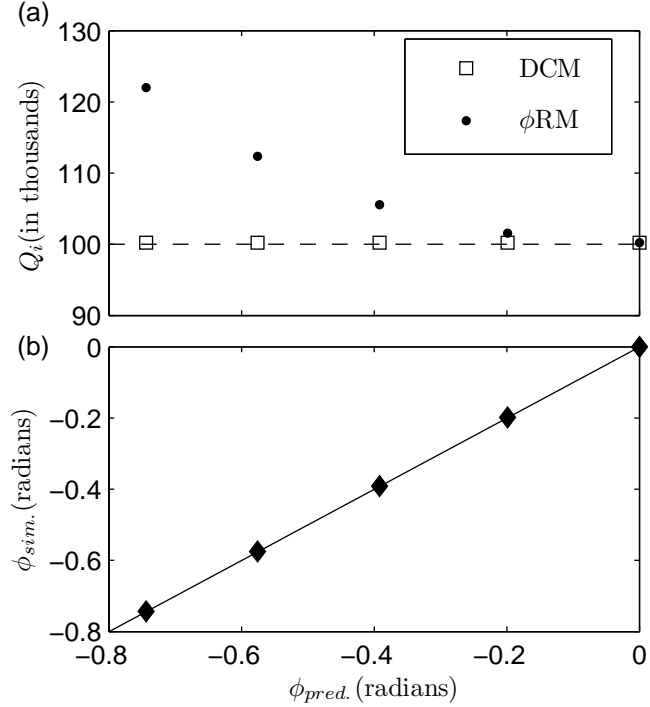}
\caption{\label{fig:Qi_phi} (a) $Q_i$ extracted from fits to circuit simulations using both analysis techniques, $\phi$RM ($\bullet$) and DCM ($\square$), as a function of predicted asymmetry angle, $\phi_{pred.}$, calculated using Eq.~(\ref{eq:phi}). The dashed line indicates actual simulation $Q_i$. At low asymmetry the two methods agree. As asymmetry is increased, the $\phi$RM extracted $Q_i$ deviates from the actual $Q_i$. (b)  The fit extracted asymmetry angle, $\phi_{sim.}$ ($\blacklozenge$), as a function of predicted asymmetry angle, $\phi_{pred.}$. The solid line is the $\phi_{pred.}=\phi_{sim.}$ line. Good agreement of that line with the results ($\blacklozenge$) indicates that this method is accurate at predicting the asymmetry.}
\end{figure}

Two internal quality factors are extracted from these fits, one using the $\phi$RM and the other using the DCM. In  Fig.~\ref{fig:Qi_phi} both extracted quality factors as well as the fit extracted asymmetry angle $\phi$ are plotted against the predicted $\phi$ from Eq.~(\ref{eq:phi}). It is clear from Fig.~\ref{fig:Qi_phi} that the DCM is more accurate as the asymmetry, $\phi$, increases, and that the two methods agree  for small asymmetry.

\begin{figure}
\includegraphics[trim = 0.2cm 0.8cm 0.2cm 0, clip]{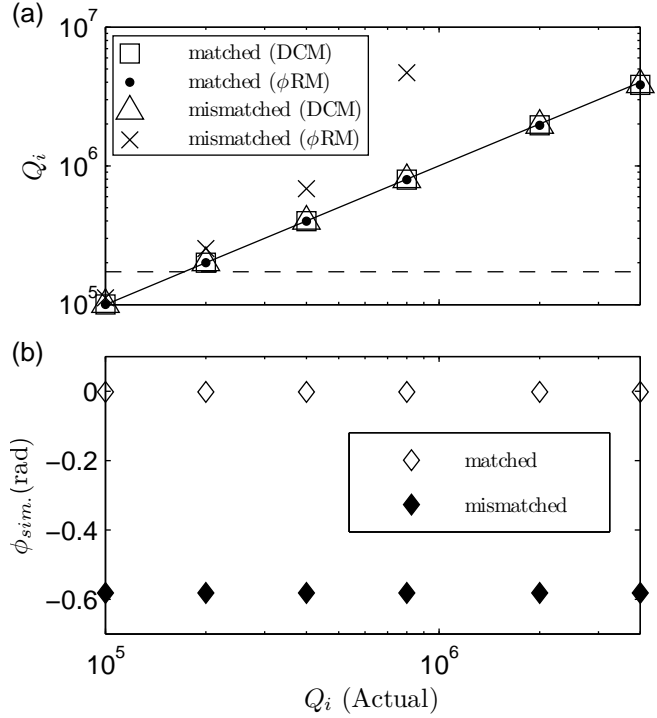}
\caption{\label{fig:Qi_sim} (a) $Q_i$ extracted with both analysis techniques ($\phi$RM and DCM) as a function of the actual $Q_i$ from two sets of simulations. The first set of simulations had high asymmetry (mismatched ports: $Z_{in}=24.5\Omega$, $Z_{out}=84.5\Omega$) and the second had low asymmetry (matched ports: $Z_{in}=Z_{out}=50\Omega$). Solid line is actual $Q_i$ equal to extracted $Q_i$ line and dashed line indicates the coupling ($Q_e$). Both analysis techniques work well with low asymmetry but only the DCM works with high asymmetry at large $Q_i$. At low simulation internal quality factors ($Q_i=10^5$) the DCM extracted internal quality factors ($Q_i=1.002\times10^5$) with less than $1\%$ deviation from the actual value in both low and high asymmetry simulations and at high simulation internal quality factors ($Q_i=4\times10^6$) the DCM extracted internal quality factors ($Q_i=3.85\times10^6$) with less than $4\%$ deviation from the actual value for both low and high asymmetry simulations. The deviation at high internal quality factors is limited numerically by the fit and is not a limit on the method. (b) The fit extracted asymmetry angle, $\phi_{sim.}$, for both low ($\lozenge$) and high ($\blacklozenge$) asymmetry simulations plotted against the actual simulation $Q_i$.}
\end{figure}

We also compared both analysis techniques when asymmetry is held constant but $Q_i$ is varied, which we will show later corresponds to some experimental data sets. Figure~\ref{fig:Qi_sim} shows the fit extracted $Q_i$ from both analysis techniques as the actual simulation $Q_i$ is increased for two sets of simulations, one with low and one with high asymmetry. For low asymmetry (matched ports) both analysis techniques yield the correct $Q_i$ within the expected first-order error ($C_C/C$). However, for high asymmetry (mismatched ports), the $\phi$RM yields quality factors that are systematically too high. For sufficiently high $Q_i$s, the $\phi$RM yields negative $Q_i$s (this is why the asymmetric data analyzed by the $\phi$RM appears to stop at large $Q_i$ in Fig.~\ref{fig:Qi_sim}). These unphysical, negative, $Q_i$s can best be understood by examining the circle plots in Fig.~\ref{fig:circle}. As discussed earlier when there is a large asymmetry, in addition to being rotated, the resonance circle grows by a factor of $1/\cos(\phi)$ (assuming full transmission off resonance). Since the $\phi$RM only rotates the circle back, it does not account for the increase in size, shown in Fig.~\ref{fig:circle}(b). So if $Q \approx Q_e$ $(Q_i>>Q_e)$, the circle diameter is larger than 1, almost crossing the y-axis. Rotating the circle using the $\phi$RM causes the circle to cross the y-axis and this yields a negative $Q_i$. In Fig.~\ref{fig:circle}(b) the $\phi$RM analyzed simulation almost crosses the origin. This corresponds to the $Q_i=8 \times 10^5$ simulation in Fig.~\ref{fig:Qi_sim}; it is an example of a simulation data set with a $Q_i$ and asymmetry not large enough to create a negative $Q_i$ but still large enough to create a considerable discrepancy between $Q_{i,\ \phi RM}$ and $Q_{i,\ DCM}$.

\begin{figure}
\subfloat{\label{fig:subfig1}\includegraphics[trim = 0.2cm 0 0 0, clip, scale=0.9]{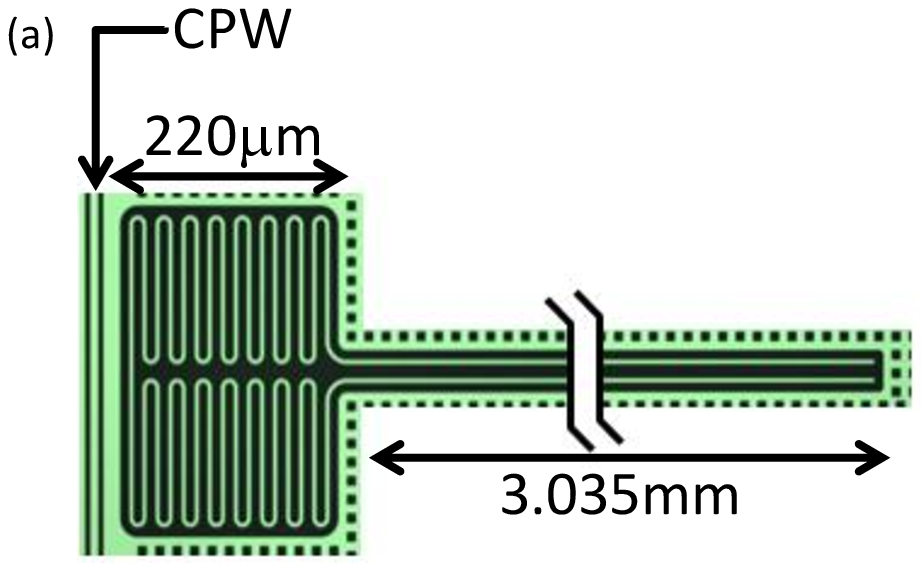}}\\
\subfloat{\label{fig:subfig2}\includegraphics[scale=1]{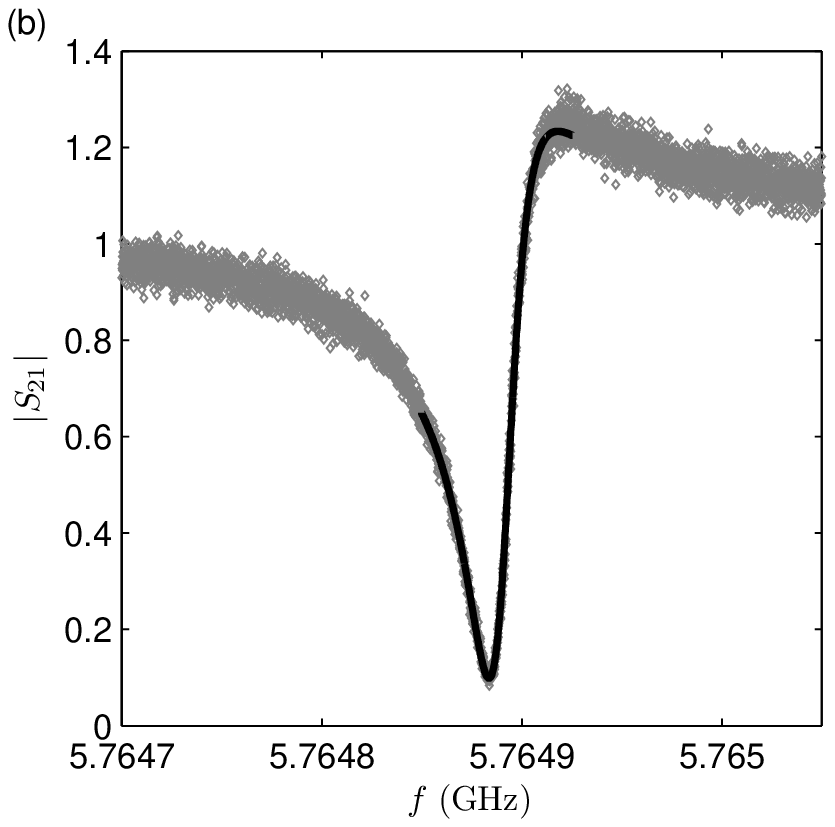}}
\caption{\label{fig:res} An image of a coplanar superconducting aluminum resonator, coupled to a coplanar waveguide (CPW) transmission line. The resonator is composed of a meandering inductor (left) and a long coplanar strip (right). In one sample box mounting this device shows high asymmetry. (b) An example of a measured line shape of the resonator with its fit. Note that the fit is centered at the resonance frequency, but not at the minimum transmission frequency because those are not the same frequencies for asymmetric line shapes.}
\end{figure}

\begin{figure}
\includegraphics[trim = 0.2cm 1.3cm 0.2cm 0, clip]{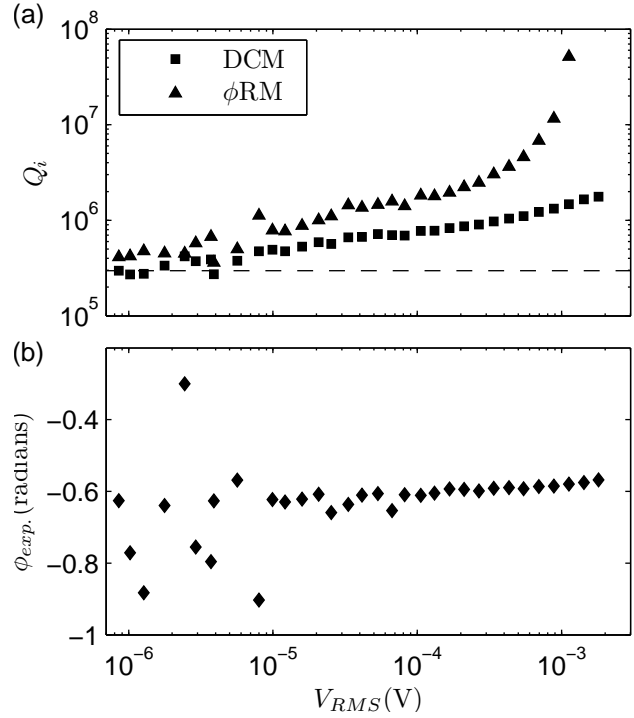}
\caption{\label{fig:Qi_data} Data from a resonator identical to that shown in Fig. 6. (a) $Q_i$, extracted using both analysis techniques, $\phi$RM ($\blacktriangle$) and DCM ($\blacksquare$), as a function of voltage across the resonator, $V_{RMS}$. As with simulated results in Fig.~\ref{fig:Qi_sim}, the $\phi$RM systematically extracts higher $Q_i$s with the highest $Q_i$s yielding negative results. The dashed line indicates the fit extracted external quality factor, $Q_e$. (b) The fit extracted asymmetry angle, $\phi_{exp.}$ ($\blacklozenge$), plotted against $V_{RMS}$. Again similar to the simulated results in Fig.~\ref{fig:Qi_sim}, $\phi_{exp.}$ is independent of the changing $Q_i$.}
\end{figure}

To further evaluate both methods, we also analyzed data from a 5.75GHz coplanar aluminum resonator on sapphire, measured at 30 mK in a dilution refrigerator. Figure~\ref{fig:res}(a) shows a picture of this resonator and a more detailed description can be found in Ref.~\onlinecite{Khalil2011}. The resonator is measured by being mounted in a copper sample box and electrical connections are made with the Coplanar waveguide (CPW) using aluminum wire bonds. Figure~\ref{fig:res}(b) shows an example of the measured resonance line shape and its fit for one mounting of the resonator which exhibited a particularly high asymmetry, presumably due to the non-ideal mounting of the device in the sample box. In Fig.~\ref{fig:Qi_data} we show the extracted $Q_i$ using both techniques and the extracted asymmetry angle, $\phi$. The quality factor dependence on voltage is discussed in Ref.~\onlinecite{Khalil2011}. Here we focus on the difference between the two analysis techniques. Figure~\ref{fig:Qi_data} is very similar to the high asymmetry simulations in Fig.~\ref{fig:Qi_sim}.  As expected from Eq.~(\ref{eq:Gd_expand}), the asymmetry, $\phi$, is independent of $Q_i$ for both the real device measurements and the simulated data. Also the last two data points for the $\phi$RM in Fig.~\ref{fig:Qi_data} are negative (and off the plot) in the same manner that the last points in the simulated data of Fig.~\ref{fig:Qi_sim} are negative.

\begin{figure}
\includegraphics[trim = 0 1cm 0.3cm 0, clip]{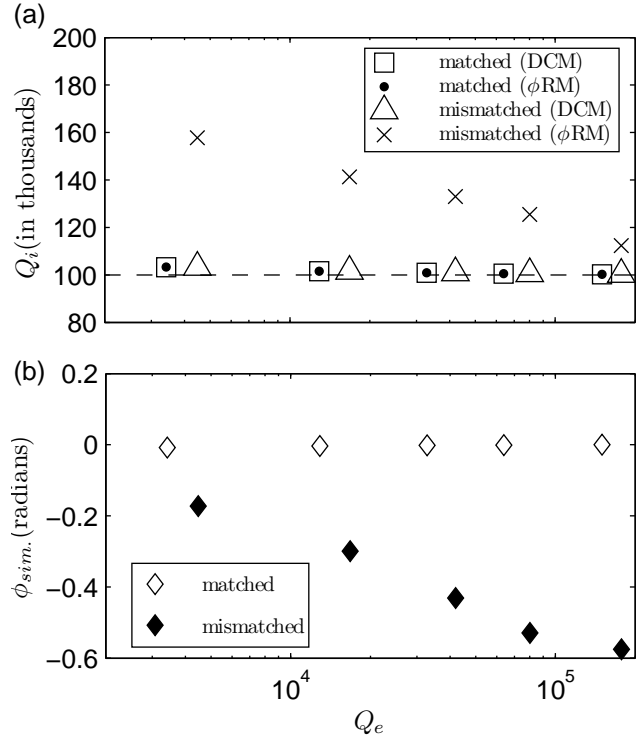}
\caption{\label{fig:Qe} (a) $Q_i$ extracted using both analysis techniques from two sets of simulations. One with high asymmetry (mismatched ports: $Z_{in}=24.5 \Omega$, $Z _{out}=84.5\Omega$) and one with low asymmetry (matched ports: $Z_{in}=Z_{out}=50\Omega$) plotted against a varying $Q_e$. $Q_e$ is varied by varying the coupling capacitance (1-10 fF), with a constant mutual inductance (5 pH). The dashed line indicates the actual $Q_i$ of the simulations. With increasing $Q_e$ the inaccuracy of the $\phi$RM is diluted due to the decreasing weight of $Q_e$ in the analysis. (b) The extracted asymmetry angles for the two simulations, low ($\lozenge$) and high ($\blacklozenge$) asymmetry.}
\end{figure}

An additional way to test the analysis techniques is by varying $Q_e$ while keeping $Q_i$ constant. In Fig.~\ref{fig:Qe}, $Q_e$ is increased by lowering the capacitive coupling. As expected, for low asymmetry (matched ports) both analysis techniques do a good job of extracting $Q_i=10^5$. However, with high asymmetry (mismatched ports) the $\phi$RM overestimates $Q_i$ by a decreasing amount as $Q_e/Q_i$ increases. Interestingly, in the $\phi$RM, as $Q_e$ increases, the extracted $Q_i$ approaches the real value, although the asymmetry, $\phi$, is increasing. This is because as $Q_e$ increases the $\phi$RM is less sensitive to the asymmetry. This behavior is captured in Eq.~(\ref{eq:Qi_right_wrong_diff}), which shows that as $Q_e$ increases, the difference between the two analysis methods vanishes. In fact for $Q_e>>Q_i$, the asymmetry becomes completely irrelevant and the two techniques converge.

\section{Conclusion}

In summary, we derived an analytical resonance line shape based on circuit parameters and found that for non-ideal conditions the line shape is asymmetric. We developed a technique (DCM) for extracting accurate internal quality factors from asymmetric resonator measurements using only transmission data. By analyzing simulated resonator measurements we found that the DCM is superior at extracting accurate internal quality factors to the conventional $\phi$RM used in millikelvin resonator measurements. We found that in the limit where the asymmetry is low, the two methods agreed, but when the asymmetry is high, particularly when $Q_i>>Q_e$, the DCM accurately determines $Q_i$ while the $\phi$RM systematically overestimates it. Also, for sufficiently high asymmetry and coupling the $\phi$RM gives a negative $Q_i$. We have also shown that the two methods can produce different results on real data taken on a coplanar superconducting aluminum resonator with high asymmetry.

\begin{acknowledgments}
We wish to acknowledge S. Anlage, C. Lobb, and S. Gladchenko for helpful discussions.
\end{acknowledgments}

%

\end{document}